\documentclass[preprint]{aastex631}

\usepackage{natbib}
\accepted{November 26, 2025}

\submitjournal{Astronomical Journal}                                                                                                                 

\begin{document}

\title{Variable X-ray Emission of the Planet Hosting T Tauri Star CI Tau}

\correspondingauthor{Stephen L. Skinner}
\email{stephen.skinner@colorado.edu}

\author[0000-0002-3025-3055]{Stephen L. Skinner}
\affiliation{Center for Astrophysics and
Space Astronomy (CASA), Univ. of Colorado,
Boulder, CO, USA 80309-0389}

\author{Manuel  G\"{u}del}
\affiliation{Dept. of Astrophysics, Univ. of Vienna,
T\"{u}rkenschanzstr. 17,  A-1180 Vienna, Austria}


\newcommand{\ltsimeq}{\raisebox{-0.6ex}{$\,\stackrel{\raisebox{-.2ex}
{$\textstyle<$}}{\sim}\,$}}
\newcommand{\gtsimeq}{\raisebox{-0.6ex}{$\,\stackrel{\raisebox{-.2ex}
{$\textstyle>$}}{\sim}\,$}}

\begin{abstract}
We report results of Chandra X-ray observations of CI Tau,
a young magnetically active classical T Tauri star 
for which previous studies have reported periodic variability attributed to 
a massive planet in a short-period orbit.
CI Tau was clearly detected
by Chandra in four separate observations acquired in late 2023.
The X-ray emission was steady in the first two observations 
with a characteristic plasma temperature kT $\approx$ 2 keV ($\approx$23 MK)
and X-ray luminosity log L$_{x}$ = 29.74 erg s$^{-1}$. 
During each of the last two observations obtained two weeks later
the count rate increased slowly 
and the X-ray plasma temperature was much higher but remained 
nearly steady at kT $\approx$ 4 - 5 keV ($\approx$46 - 58 MK)
and peak luminosity log L$_{x}$ = 30.5 erg s$^{-1}$.
Such variable X-ray emission in T Tauri stars accompanied by high plasma 
temperatures is a signature of 
magnetic activity, consistent with the known presence of a strong
magnetic field in CI Tau. We summarize the variable X-ray emission 
properties of CI Tau within the framework of T Tauri stars of
similar mid-K spectral type, identify possible variability mechanisms,
and assess the effects of stellar X-ray irradiation on the claimed planet.
\end{abstract}


\section{Introduction}
Periodic radial velocity (RV) variations in the 
$\sim$2 Myr old classical T Tauri star (cTTS) CI Tau 
observed by Johns-Krull et al. (2016) were attributed to
a massive planet in a $\approx$9 day orbit (CI Tau b). 
Additional spectropolarimetric data acquired by
Donati et al. (2020; 2024)
confirmed RV periodicity with
a period of 9.01$\pm$0.023 d which they interpreted as
the stellar rotation period, not a planet signature.
Donati et al. (2020; 2024) showed that CI Tau has a strong
magnetic field whose radial component reaches 
$\approx$3.7 kG in a dark photospheric spot.
A mean magnetic field strength B$_{avg}$ $\approx$ 2.2 kG
was also determined from analysis of high-resolution near-IR
spectra by Sokal et al. (2020).

More recent photometric and spectroscopic analysis
by Manick et al. (2024) using K2 and ground-based
data detected multiple coherent periods of 
approximately 6.6, 9, 11.5, 14.2, and 25.2 days.
The 25.2 day period is the most consistent and stable
and they conclude it is best interpreted as the 
signature of a 3.6$\pm$0.3 M$_{\rm Jup}$ planet
(CI Tau c) in an eccentric orbit at a semi-major axis
$a$ = 0.17$\pm$0.08 au. 
But Donati et al. (2024) attributed the $\approx$25 d
period to non-axisymmeteric structure in the inner disk.
If the $\approx$25 d period is  
indeed a planet signature then CI Tau c would be an
important and rare example of a Jupiter-mass
planet in a tight orbit around a young cTTS.

T Tauri stars typically exhibit strong variable 
X-ray emission including powerful flares that is 
usually interpreted as thermal emission from hot 
magnetospheric plasma at temperatures T $\sim$ 10$^{6}$ - 10$^{8}$ K
confined in magnetic structures, analogous to the solar corona
(e.g. Favata et al. 2005; Wolk et al. 2005).
In a few cases, softer X-ray emission has been detected in cTTS
and attributed to accretion streams with TW Hya being a
well-known example (e.g. Kastner et al. 2002; Stelzer \& Schmitt 2004).
 
X-ray emission from CI Tau will irradiate and heat the 
giant planet's atmosphere and will need to
be taken into account in realistic atmospheric models.
We present here results of the first Chandra pointed 
observations of CI Tau which show that its X-ray emission
is strongly variable. We summarize its X-ray variability and
spectral properties and estimate the ionization, heating, and
mass-loss rates of the planet due to star's incident high-energy
radiation.

\section{X-ray Observations}

\subsection{Chandra}

Chandra observed CI Tau in late 2023 using the 
Advanced CCD Imaging Spectrometer 
(ACIS-S)\footnote{Detailed information on ACIS-S is given in
ch. 6 of the Chandra Proposer's Observatory Guide:~
https://cxc.harvard.edu/proposer/POG/}.
ACIS-S provides coverage in the E $\approx$ 0.3-10 keV
energy range, but the effective area below $\approx$1 keV
is now low due to cumulative contaminant buildup during the mission.
ACIS-S has a native pixel size of 0$''$.49 and on-axis 
spatial resolution of $\approx$0$''$.5. 
The undispersed ACIS-S CCD spectra have energy resolution  
of $\approx$100 eV (FWHM @1.5 keV).
Data were reduced using the Chandra Interactive Analysis of 
Observations (CIAO v. 4.14) software package and recent
calibration data. The total 65 ks observing time
was split into four observations (ObsIds) due to Chandra's
operational constraints as summarized in Table 1.
Source events and spectra were extracted from a circular region of
radius 2$''$ centered on the X-ray peak. Background was negligible.
Spectra were fitted and analyzed using XSPEC v 12.10.1.

\subsection{XMM-Newton}

CI Tau was also detected as an X-ray source in the XMM-Newton Extended Survey
of Taurus (XEST) in 2005 (ObsId 0203541701) with a usable exposure 
of 27.575 ks and is identified as XEST no. 17-058 in the catalog of
G\"{u}del et al. (2007). The XMM-Newton data provide a useful comparison
with the more recent Chandra data.
The XEST catalog summarizes a two-temperature (2T) thermal plasma model 
fit of the XMM-Newton spectra using variable abundances which converged to
an equivalent neutral hydrogen column density 
N$_{\rm H}$ = 0.62 (0.31 - 1.10) $\times$ 10$^{22}$ cm$^{-2}$,
cool and hot plasma component temperatures 
T$_{1}$ = 7.88 MK (kT$_{1}$ = 0.68 keV) and
T$_{2}$ = 43.71 MK (kT$_{2}$ = 3.77 keV), 
ratio of cool to hot component 
emission measures EM$_{1}$/EM$_{2}$ = 0.94/1.65,  
T$_{avg}$ = 23.44 MK (kT$_{avg}$ = 2.02 keV), and X-ray luminosity
log L$_{x}$(0.3 - 10 keV) = 29.64 erg s$^{-1}$ when normalized to the 
Gaia DR3 parallax distance of 160.3 pc. 
The above value of T$_{avg}$  is the EM-weighted logarithmically-averaged temperature 
log T$_{avg}$ = (EM1$\cdot$log T$_{1}$ $+$ EM2$\cdot$log T$_{2}$)/(EM1 $+$ EM2).    
Further analysis of
the XMM-Newton data by Stelzer et al. (2007) reported flare-like variability
of duration $>$5.5 ks and a flare luminosity log L$_{x,F}$ = 29.82 erg s$^{-1}$,
normalized to 160.3 pc. The Chandra observations discussed below reveal
X-ray luminosities a factor of $\approx$5 higher. 

\vspace*{0.6in}


\begin{deluxetable}{lllll}
\tabletypesize{\small}
\tablewidth{0pt}
\tablecaption{Summary of CI Tau Chandra Observations\tablenotemark{a}}
\tablehead{
\colhead{Parameter} &
\colhead{}          &
\colhead{Observation}          &
\colhead{}          \\
}
\startdata
ObsId (state)                    &  28364 (low)        & 29094 (low)      & 28501 (high)    & 29122 (high)         \\
Start Date (2023)/Time (TT)      &  Nov. 27/00:57      & Nov. 28/13:07    &  Dec. 12/02:06  & Dec. 12/18:57  \\
Stop  Date (2023)/Time (TT)      &  Nov. 27/07:08      & Nov. 28/18:16    &  Dec. 12/06:01  & Dec. 12/23:59  \\
Livetime (ks)\tablenotemark{b}   & 19.319           & 16.360           &  11.921         & 15.869      \\
Counts (0.3-8 keV)               & 151                 & 128              &  458            & 678        \\
Rate (cts/ks)\tablenotemark{b}   & 7.82                & 7.82             &  38.42 (v)      & 42.72 (v)  \\
\enddata
\tablenotetext{a}{Data were obtained using ACIS-S in faint timed event mode,
a frame time of 3.0 s, and the source positioned on CCD S3. 
The X-ray centroid position of CI Tau is (J2000)
R.A. = 04$^{\rm h}$33$^{\rm m}$52$^{\rm s}$.04, 
decl. = $+$22$^{\circ}$50$'$29$''$.5.
The {\em Gaia} DR3 position is R.A. = 04$^{\rm h}$33$^{\rm m}$52$^{\rm s}$.02,
decl. = $+$22$^{\circ}$50$'$29$''$.8. 
}
\tablenotetext{b}{Livetime excludes operational and instrumental overheads such as CCD readout times.
Rate = counts/livetime. A (v) denotes variable count rate.}
\end{deluxetable}


\section{RESULTS }

\subsection{X-ray Light Curves and Time Variability}

The X-ray emission of 
CI Tau  was steady in the first two Chandra observations
on 27-28 Nov. 2023 with a mean broad-band count rate of 
7.82$\pm$3.15 c ks$^{-1}$ (0.3 - 8 keV). We refer to 
this as the {\em low state} count rate. 
But as seen in Table 1 and Figure 1, the count rate 
in the final two observations (ObsIs 28501 and 29122) 
two weeks later was significantly higher and variable with 
a mean rate of 40.5$\pm$12.0 c ks$^{-1}$ (0.3 - 8 keV). 
We refer to these two observations as {\em high state}. 
The emission was harder during the two high state
observations as gauged by the hardness ratio 
H.R. = counts(2-8 keV)/counts(0.3-8) keV 
and the mean plasma temperature determined from spectral
fits (Sec. 3.2) was also higher, as summarized in Table 2.
Even though the count rate was generally increasing in the
final two observations the H.R. showed
no statistically significant change during either observation
(bottom panel of Fig. 1-top).

The broad-band light curve in ObsId 28501 shows a 
gradual increase during the $\approx$3 hour observation
with no clear sign of a turnover. Thus, the count rate
may have continued to increase after ObsId 28501 was
terminated. But $\approx$13 hours later when the final
observation ObsId 29122 began the count rate had decreased
back to nearly the low state value. So the decay 
timescale of ObsId 28501 is broadly constrained to $\leq$13 hours.

The light curve of ObsId 29122 is similar to that of
ObsId 28501 except that the broad-band count rate does 
show a peak and turnover near the end of the observation. 
To estimate the rise time we fitted the ObsId 29122 light 
curve with a simple sinusoid and extrapolated the fit 
backward to its point of intersection with the low state
count rate (Fig. 1-bottom). Assuming that the low state
value represents the typical non-flaring count rate of
CI Tau, a rise time of $\approx$4 hours is inferred for 
ObsId 29122. However, the decay time is not constrained 
for ObsId 29122 since the observation terminated shortly
after the light curve peaked and turned over.
Obviously, additional uninterrupted
time monitoring is needed to better characterize the light curve
variability profiles and timescales.

\begin{figure}
\figurenum{1}
\includegraphics*[width=7.0cm,height=9.8cm,angle=-90]{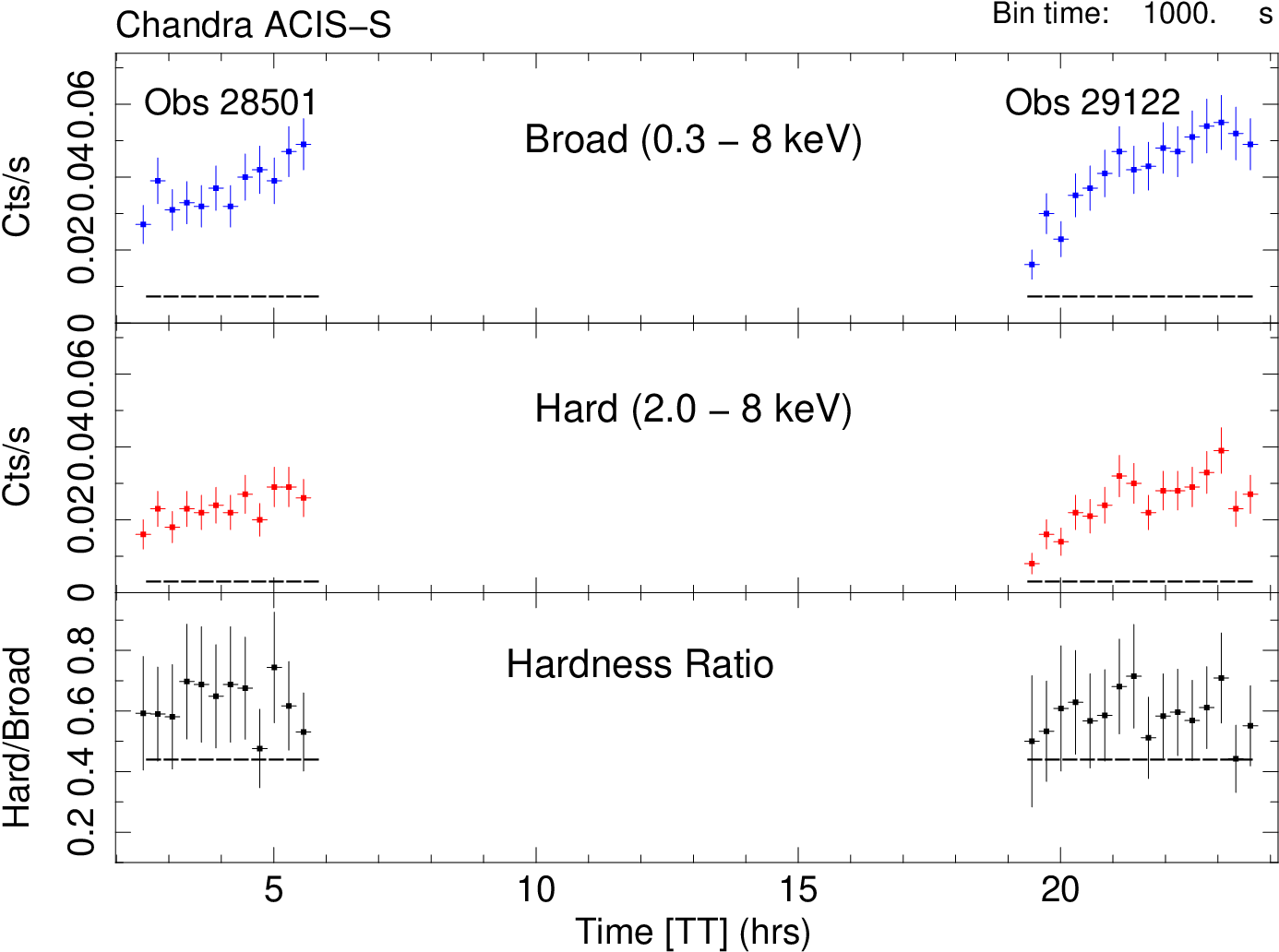} \\ \\
\includegraphics*[width=7.0cm,height=9.8cm,angle=-90]{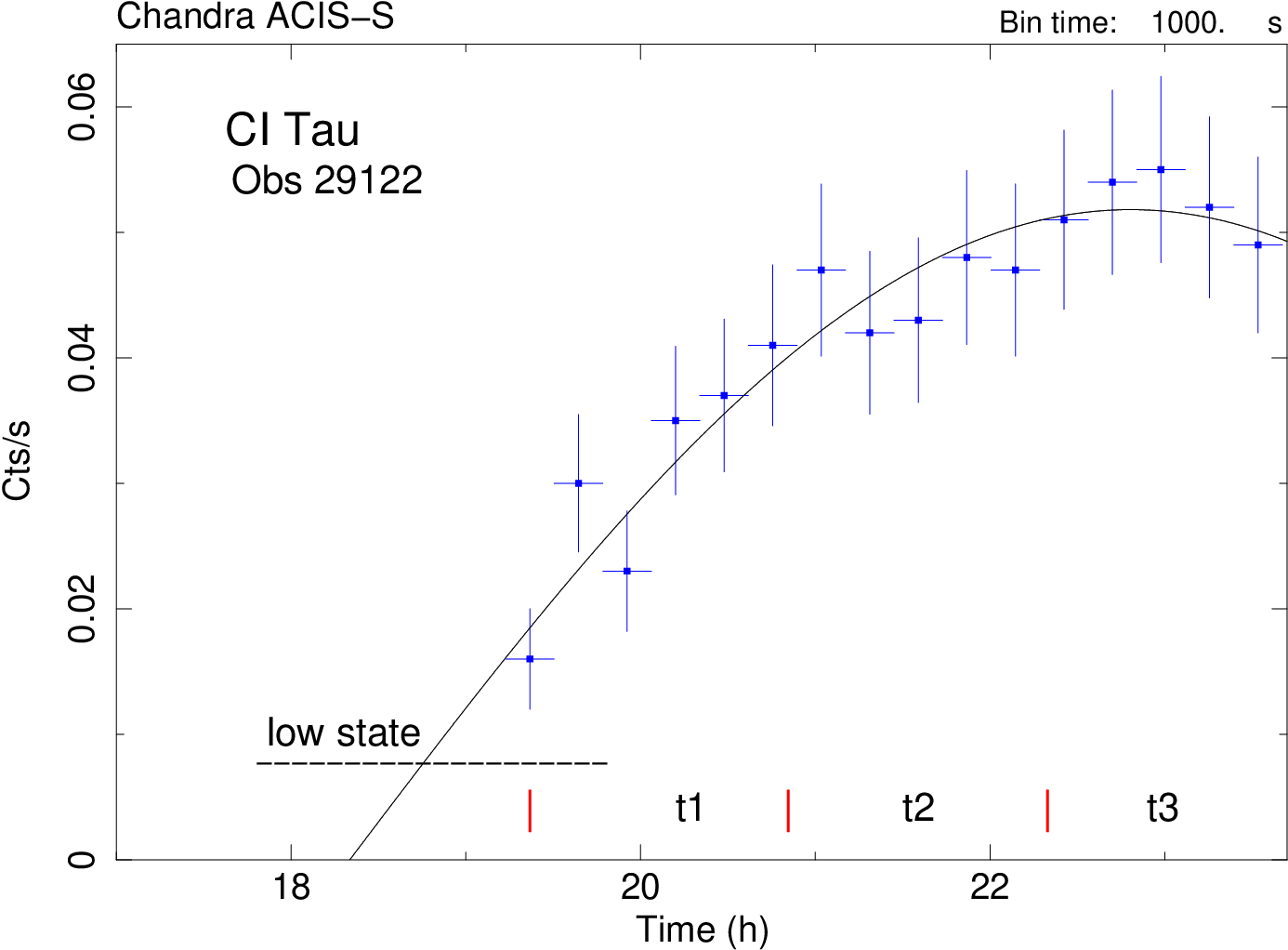}
\caption{Chandra ACIS-S light curves of CI Tau binned to 1000 s intervals.
Error bars are 1$\sigma$.
Terrestrial Time (TT) (hours) is on 12 Dec 2023.
{\bf Top:} Broad (0.3 - 8 keV) and hard (2 - 8 keV) energy bands 
and the hard-band/broad-band count rate ratio for ObsIds 28501 and 29122. 
For comparison, the dashed lines show the corresponding values
for the initial observations two weeks earlier when CI Tau showed fainter 
non-variable emission (low state).
{\bf Bottom:} A sinusoid fit (solid line) of the broad-band light curve of CI Tau for ObsId 29122 showing
a slow turnover toward end of the observation. Extrapolating the fit back to the low state 
broad-band count rate of 7.82 c/ks gives an inferred rise time to peak count rate 
of $\approx$4 hours. 
The three intervals use to extract the time-partitioned spectra shown in Fig. 2-bottom are marked.
}
\end{figure}

\subsection{X-ray Spectra}

The spectra for all four observations are overlaid in Figure 2-top.
We compared fits of the Chandra spectra using absorbed one-temperature (1T) 
and two-temperature (2T) apec optically thin plasma models in XSPEC.
No significant fit improvement was found using the 2T models and the
1T model fits are summarized in Table 2. The X-ray absorption column
density N$_{\rm H}$ is not tightly constrained by the ACIS-S spectra
since absorption mainly affects the spectrum below $\approx$1 keV 
where the ACIS-S sensitivity (effective area) is now quite low.
We thus compared ACIS-S spectral fits allowing N$_{\rm H}$ to vary 
versus holding it fixed at the previously determined XMM-Newton value
N$_{\rm H}$ = 0.62 (0.31 - 1.10; 1$\sigma$) $\times$ 10$^{22}$ cm$^{-2}$.
XMM-Newton provides better sensitivity
below 1 keV than Chandra ACIS-S and thus a more reliable N$_{\rm H}$ determination. 
The above value of N$_{\rm H}$ corresponds to
A$_{\rm V}$ = 3.2 (1.6 - 5.8) mag using the conversion
N$_{\rm H}$ (cm$^{-2}$)= 1.9$\pm$0.3$\times$10$^{21}$A$_{\rm V}$
(Gorenstein 1975; Vuong et al . 2003). This X-ray derived
extinction is larger than the value A$_{\rm V}$ = 2.6$\pm$0.2 mag 
obtained by Gangi et al. (2022) but the two values are consistent
to within the uncertainties.

Since the ACIS-S count rate was nearly constant in the first
two observations (ObsIds 28364 and 29094, low state) their spectra were 
fitted simultaneously to better constrain fit parameters.
The fits with either N$_{\rm H}$ held fixed or allowed to vary give
similar results with a best-fit temperature kT = 2.25$^{+0.73}_{-0.59}$ keV
and log L$_{x}$ = 29.74$\pm$0.3 erg s$^{-1}$ at d = 160.3 pc.
The values of kT and L$_{x}$ are nearly the same as
obtained by XMM-Newton in 2005 (Sec. 2.2). 
If  N$_{\rm H}$ is allowed to vary it converges to
a value slightly larger than, but consistent with,
the value obtained by XMM-Newton.

Spectra for the last two observations (ObsIds 28501 and 29122, high state)
when the count rate was variable were fitted separately.
The plasma temperature also varied during ObsIds 28501 and 29122
(discussed below) so spectral fits based on all events detected
for each observation are time averages.
These fits (Table 2) yield plasma temperatures kT = 5$\pm$1.5 keV
with higher temperatures inferred when N$_{\rm H}$ was held fixed
at the XMM-Newton value. Lower temperatures and slightly better fits
as judged by  $\chi^2$ statistics were obtained by varying N$_{\rm H}$.
When N$_{\rm H}$ is varied the best-fit values are slightly higher
than than obtained for the low state spectra but their 1$\sigma$
confidence ranges overlap so any change in N$_{\rm H}$ between
low and high state is of low significance. But a comparison of
the high state spectral fits using either fixed or variable N$_{\rm H}$
clearly shows that the mean X-ray temperature kT, observed (absorbed)
X-ray flux F$_{x,abs}$, and luminosity L$_{x}$ were $\approx$5 - 6 times
higher than during the low state observations two weeks earlier.
Using the fit results of CI Tau in Table 2 we adopt characteristic
X-ray luminosities log L$_{x}$ (erg s$^{-1}$) = 29.74 (low state, 
ObsIds 28364 and 29094) and 30.50 (high state, ObsIds 28501 and 29122).
Adopting the CI Tau stellar luminosity 
log(L$_{*}$/L$_{\odot}$) = 0.1 from Donati et al. (2020) gives
log(L$_{x}$/L$_{*}$) = $-$3.94 (low state) and $-$3.18 (high state).

Very few emission lines can be positively identified in the spectra.
But the high state spectra  of ObsId 28501 and 29122 (Fig. 2)
show a broad feature spanning the energy of the Si XIII triplet 
(1.839 - 1.865 keV). There are also narrow peaks  near 2.46 and 2.88 keV 
that may be S XV emission, and possible emission from Ar XVII (3.32 keV).
There is no line emission from the Fe XXV
complex at 6.64 keV but some emission is visible in the ObsId 29122
spectrum up to 6.4 keV.

\begin{figure}
\figurenum{2}
\includegraphics*[width=7.0cm,height=9.8cm,angle=-90]{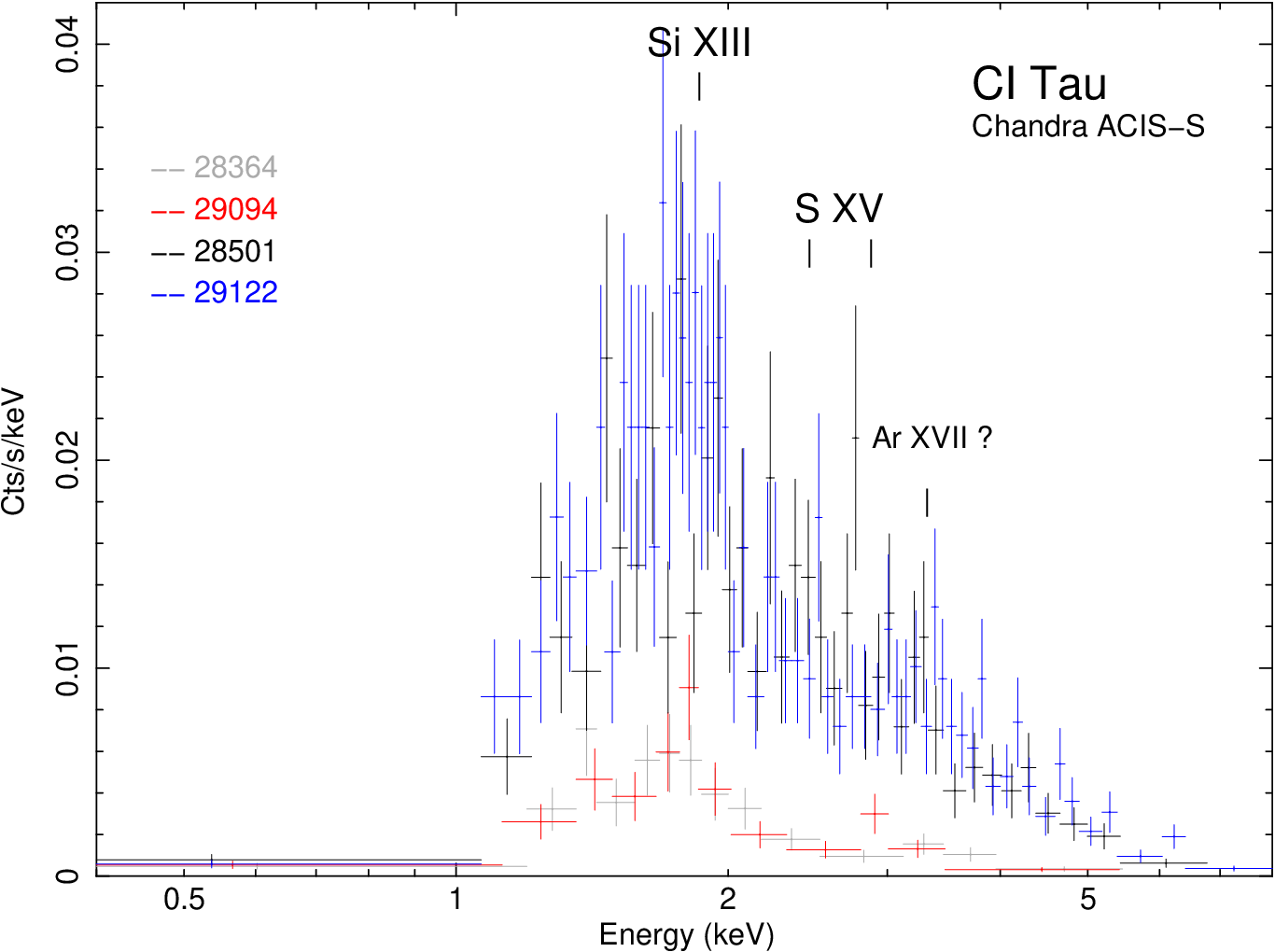} \\
\includegraphics*[width=7.0cm,height=9.8cm,angle=-90]{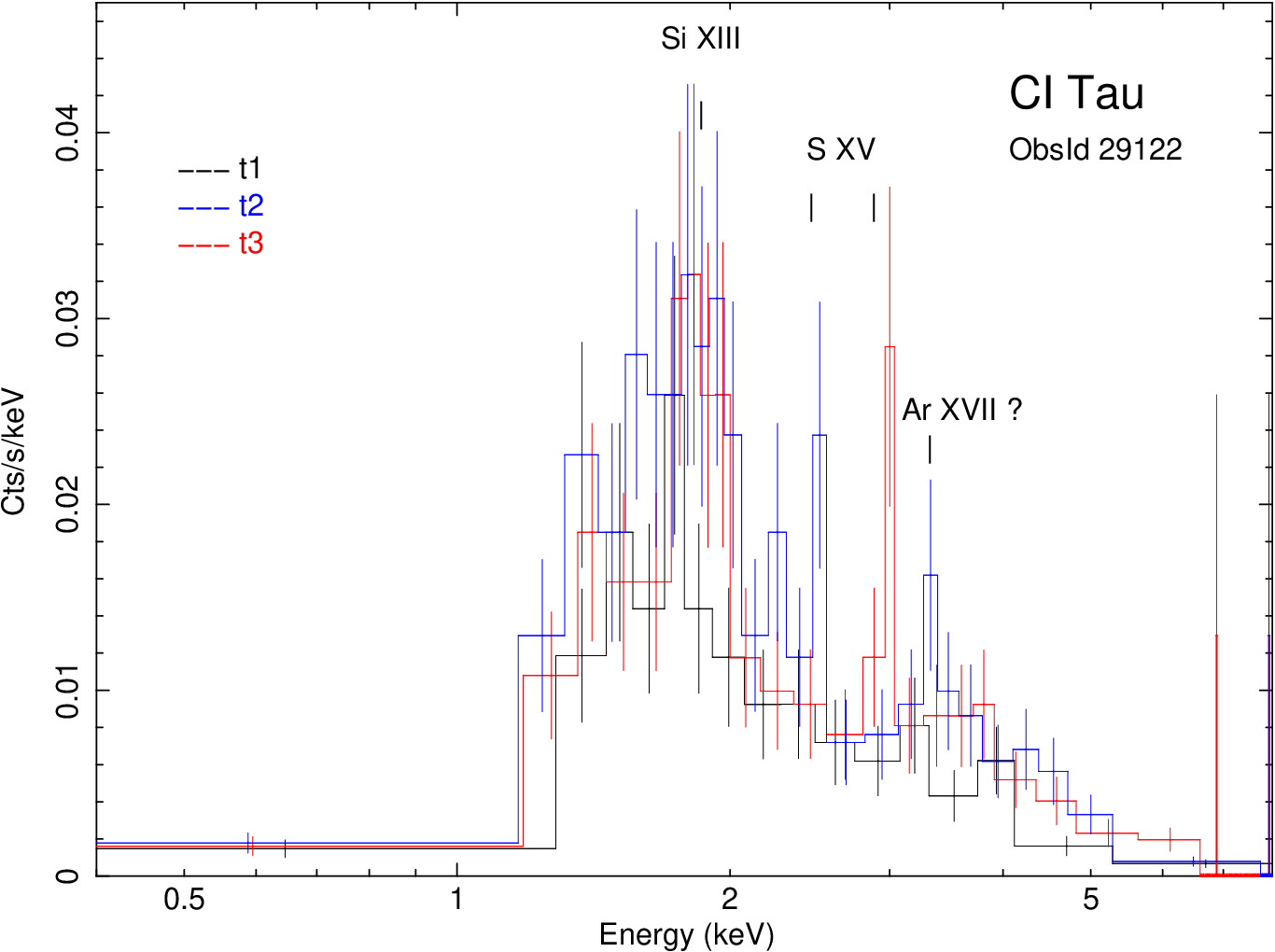} \\
\caption{Chandra ACIS-S spectra of CI Tau binned to a minimum of 10 cts per bin.
Error bars are 1$\sigma$.
{\bf Top:} Overlay of spectra from all four observations.
{\bf Bottom:} Overlay of the spectra for ObsId 29122 obtained by partitioning
the data into three equal time intervals: first (t1), middle (t2), and last (t3).
Note enhanced line emission during intervals t2 and t3 from Si XIII and possible
faint S and Ar lines.
}
\end{figure}

\begin{deluxetable}{lcclcccl}
\tabletypesize{\scriptsize}
\tablewidth{0pt}
\tablecaption{Summary of CI Tau X-ray Spectral Properties\tablenotemark{\rm \small{a}}}
\tablehead{
           \colhead{ObsId\tablenotemark{\rm \small{b}}}      &
           \colhead{Hardness}           &
           \colhead{N$_{\rm H}$}        &
           \colhead{kT}                 &
           \colhead{norm}                 &
           \colhead{F$_{x,abs}$}        &
           \colhead{log L$_{x}$}        &
           \colhead{$\chi^2$/dof}        \\
           \colhead{}                   &
           \colhead{Ratio}              &
           \colhead{(10$^{22}$ cm$^{-2}$)}  &
           \colhead{(keV)}              &
           \colhead{(10$^{-4}$ cm$^{-5}$)}              &
           \colhead{(erg cm$^{-2}$ s$^{-1}$)}  &
           \colhead{(erg s$^{-1}$)}     &
           \colhead{} }
\startdata
1 \& 2            & 0.44$\pm$0.09   & [0.62]\tablenotemark{\rm \small{c}} &   2.32 (2.07 - 2.64)   & 1.11  & 9.03 (8.33 - 9.79)e-14 & 29.74     & 10.7/11 (0.98) \\
1 \& 2            & 0.44$\pm$0.09   & 0.67 (0.43 - 0.98)    &   2.18 (1.66 - 2.98)   & 1.16  & 8.84 (7.40 - 9.49)e-14 & 29.75     & 10.7/10 (1.07)  \\
                  &                 &                       &                        &       &                        &           &                \\
3                 & 0.63$\pm$0.07   & [0.62]\tablenotemark{\rm \small{c}} &   6.51 (5.33 - 8.07)   & 4.89  & 6.30 (5.95 - 6.63)e-13 & 30.45     & 18.6/19 (0.98) \\
3                 & 0.63$\pm$0.07   & 1.06 (0.83 - 1.26)    &   3.44 (2.83 - 4.68)   & 6.57  & 5.76 (4.81 - 6.03)e-13 & 30.53     & 14.8/18 (0.82) \\
                  &                 &                       &                        &       &                        &           &                \\
4                 & 0.59$\pm$0.05   & [0.62]\tablenotemark{\rm \small{c}} &   5.73 (4.86 - 6.71)   & 5.23  & 6.52 (6.12 - 6.74)e-13 & 30.48     & 27.1/30 (0.90) \\
4                 & 0.59$\pm$0.05   & 0.86 (0.71 - 1.02)    &   3.98 (3.32 - 4.90)   & 6.21  & 6.22 (5.73 - 6.41)e-13 & 30.52     & 24.6/29 (0.85) \\
                    &               &                     &                        &       &                        &                          \\
\multicolumn{8}{c}{Time Partitioned Data\tablenotemark{\rm \small{d}}}             \\
3 (t1)            & 0.62$\pm$0.11   & [1.06]             &   3.34 (2.59 - 4.77)    & 5.39   & 4.63 (3.78 - 5.10)e-13  & 30.45  & 7.27/6 (1.21) \\
3 (t2)            & 0.66$\pm$0.11   & [1.06]             &   4.05 (3.16 - 5.67)    & 5.54   & 5.32 (4.48 - 5.79)e-13  & 30.48  & 5.63/7 (0.81) \\
3 (t3)            & 0.60$\pm$0.09   & [1.06]             &   4.00 (3.13 - 5.64)    & 6.88   & 6.56 (5.35 - 7.32)e-13  & 30.57  & 13.2/9 (1.47) \\
                  &                 &                    &                         &        &                         &        &                \\
4 (t1)            & 0.58$\pm$0.10   & [0.86]             &   4.11 (3.38 - 5.18)    & 4.35   & 4.44 (3.89 - 4.73)e-13  & 30.37  & 2.51/9 (0.28) \\
4 (t2)            & 0.61$\pm$0.09   & [0.86]             &   5.74 (4.64 - 7.79)    & 5.82   & 6.82 (5.76 - 7.30)e-13  & 30.52  & 8.18/13 (0.63) \\
4 (t3)            & 0.58$\pm$0.08   & [0.86]             &   4.38 (3.76 - 5.35)    & 6.91   & 7.28 (6.51 - 7.64)e-13  & 30.58  & 13.4/15 (0.89) \\
\enddata
\tablenotetext{a}{
Notes: 
Hardness Ratio (H.R.) = counts(2-8 keV)/counts(0.3-8 keV).
Spectral fits were used to determine the
equivalent neutral hydrogen absorption column density N$_{\rm H}$
(held fixed during fitting if enclosed in brackets),
plasma temperature kT in energy units,
norm (cm$^{-5}$) = 10$^{-14}$EM/4$\pi$d$_{cm}^{2}$ where
EM is the volume emission measure and d$_{cm}$ (cm) is the distance to
the star, absorbed X-ray flux F$_{x,abs}$(0.3-8 keV),
and intrinsic (unabsorbed) X-ray luminosity L$_{x}$(0.3-8 keV) at d = 160.3 pc.
Parentheses enclose 1$\sigma$ confidence ranges.
Spectra were binned to a minimum of 20 cts/bin and fitted in XSPEC using
an absorbed one-temperature (1T) thermal plasma model of the
form N$_{\rm H}$*kT with solar metallicity $Z$= 1.0.
Absorption was modeled using the $tbabs$ model and
plasma temperature using the {\em apec} model.}
\tablenotetext{b}{ObsIds: (1) 28364 (2) 29094 (3) 28501 (4) 29122. 
ObsIds 1 \& 2 denotes simultaneous fits of spectra extracted 
using all events from both observations.}
\tablenotetext{c}{N$_{\rm H}$ held fixed at the value determined by 
XMM-Newton (G\"{u}del et al. 2007).}
\tablenotetext{d}{The time-partitioned fits for observations 3 and 4 
are based on spectra extracted from equal time intervals spanning the 
first (t1), second (t2), and last (t3) thirds of the observations
and rebinnned to a mininum of 15 cts/bin. Each of the separate time intervals for
ObsId 28501 span 4.027 ks and have (135,142,181) counts and those for ObsId 29122
span 5.361 ks and have (165,239,274) counts (0.3-8 keV).
The tripartite spectra were fitted with the same
1T apec model as for those based on the full observations
N$_{\rm H}$ was held fixed at the values determined from fits of the 
spectra extracted for the full observations.}
\end{deluxetable}

\subsubsection{Time-Partitioned X-ray Spectra}

To further characterize the variability in ObsIds 28501 and 29122
we partitioned the data for each observation into three time segments
of equal duration: first (t1), middle (t2), and last (t3).
Separate source event lists and spectra were extracted for 
each time segment and fitted with 1T apec models. 
The time-partitioned spectra contain fewer events than
the full-observation spectra (Table 2 notes) so the absorption and 
temperature are not as tightly constrained. 
To reduce the number of free parameters in the model we fixed the 
absorption at the value N$_{\rm H}$ = 0.62 $\times$ 10$^{22}$ cm$^{-2}$ 
determined by XMM-Newton and, for comparison, held it fixed at the 
slightly higher values determined from fits of the full Chandra spectra.

The mean temperatures kT determined from fits of the tripartite spectra
are sensitive to the value of N$_{\rm H}$ with the lower value from XMM-Newton
resulting in higher kT, as was also found for fits of the full spectra. 
Better fits as judged by the $\chi^2$ statistic were
obtained using the slightly higher N$_{\rm H}$ values from Chandra
and those fits are summarized in Table 2.

Fits of the tripartite spectra for ObsId 28501 suggest a
modest temperature increase occurred between the first and second
segments but the  uncertainties in kT are large enough
to accommodate a steady temperature. The best-fit
temperatures of the middle and last segments are nearly
identical so there is no indication that the plasma cooled.
But it is clear that the X-ray flux F$_{x,abs}$ and 
luminosity L$_{x}$ are higher in the last segment of ObsId 28501
than the first.

The best-fit temperatures of the tripartite spectra for 
ObsId 29122 are nearly the same for the first and last segments
but higher for the middle segment. However, the apparent
temperature increase  for the middle 
segment is of low statistical significance (1.1 - 1.3$\sigma)$ 
so a steady temperature cannot be ruled out for ObsId 29122.
Even so, some spectral changes did occur during ObsId 29122 as shown in 
the spectral overlay in Figure 2-bottom. The last two segments (t2,t3)
have enhanced emission near 1.86 keV corresponding to the
Si XIII triplet (E$_{lab}$ = 1.839 - 1.865 keV) which
emits maximum line power at T$_{max}$ $\approx$ 10 MK.
In addition, narrow peaks appear that may be faint emission lines
of S XV (E$_{lab}$ = 2.46 and 2.88 keV; T$_{max}$ $\approx$ 16 MK)
and Ar XVII (E$_{lab}$ = 3.32 keV; T$_{max}$ $\approx$ 40 MK).
Both F$_{x,abs}$ and L$_{x}$ increased between the first and
last segments of ObsId 29122, as also found for ObsId 28501.

\subsection{X-ray Emission Measure}

The XSPEC norm value returned from apec spectral fits (Table 2) is
related to the volume emission measure (EM) of the plasma by
the relation norm (cm$^{-5}$) = 10$^{-14}$EM/4$\pi$d$_{cm}^{2}$
where d$_{cm}$ is the distance to the star in cm and 
EM (cm$^{-3}$) = $\int$n$_{e}$n$_{\rm H}$dV.  
Here, n$_{e}$ and n$_{\rm H}$
are the electron and hydrogen number densities (cm$^{-3}$)
and the integration is over the volume V of the emitting plasma.
At the high X-ray temperatures T $\gtsimeq$ 10$^{7}$ K  of
CI Tau the plasma is assumed to be nearly fully-ionized and 
in that case n$_{e}$ $\sim$ n$_{\rm H}$ if the plasma is H-dominated.
If the particle density is nearly uniform in the emitting region
EM $\sim$ n$_{e}^2$V. At a distance of 160.3 pc 
one obtains EM/norm  = 3.07$\times$10$^{56}$ $\sim$ n$_{e}^2$V (cm$^{-3}$). 
To proceed further an {\em a priori} estimate of n$_{e}$ or V is needed
and we assume a plausible range of n$_{e}$ values below.

Previous studies of T Tauri stars have obtained a wide range of
electron densities 
10$^{8}$ $\ltsimeq$ n$_{e}$ $\ltsimeq$ 10$^{12}$ cm$^{-3}$
(e.g. Telleschi et al. 2007; Bary et al. 2008; Raassen 2009;
L\'{o}pez-Mart\'{i}nez \& G\'{o}mez de Castro 2014).
The lower end of this range is usually associated with 
magnetospheric plasma and the high end with accretion streams.

For CI Tau in low state the XSPEC fits assuming solar metallicity
give norm $\approx$ 1.1 $\times$ 10$^{-4}$ cm$^{-5}$. Assuming a 
magnetospheric density n$_{e}$ $\sim$ 10$^{9}$ cm$^{-3}$
yields V $\sim$ 3 $\times$ 10$^{34}$ cm$^{3}$, which 
is comparable to the stellar volume V$_{*}$ = 1.1 $\times$ 10$^{34}$ cm$^{3}$
using R$_{*}$ = 2 R$_{\odot}$.
For higher densities n$_{e}$ $\sim$ 10$^{12}$ cm$^{-3}$
characteristic of accretion streams the inferred volume is
V $\sim$ 3 $\times$ 10$^{28}$ cm$^{3}$, a small fraction of V$_{*}$.
In high state the norm is a factor of $\approx$5 - 6 greater than in
low state and the inferred value of V scales up accordingly.

\section{X-ray Variability of CI Tau in Context With Other T Tauri Stars}

X-ray variability (or ``flares'') in TTS  can be broadly characterized 
by luminosity, duration, and light curve shape as discussed by Wolk et al. (2005). 
They analyzed 41 X-ray flares detected in a sample of 28 solar-mass 
pre-main sequence stars in the Orion Nebula Cluster (ONC) observed during 
the 13.2 day Chandra Orion Ultradeep Project (COUP).
Their sample focused on stars of K5-7 spectral type, a good match for
the mid-K spectral type of CI Tau (Herczeg \& Hillenbrand 2014).

X-ray luminosities in the K5-7 sample during periods of elevated
count rate above the characteristic non-flaring rate
were log L$_{x}$ = 29.8 - 31.2 ergs s$^{-1}$
which encompasses the high state value log L$_{x}$ = 30.5 ergs s$^{-1}$
observed here for CI Tau.
Flare durations were 1 hr to 3 days, a large range which 
accommodates the variability of CI Tau reported here,
albeit based on only four short observations of $\ltsimeq$5 hours each.
Hot-component temperatures determined from 2T thermal model
spectral fits of the ONC K5-7 stars were in the range 
$\approx$2 - 6 keV but much higher in a few cases. 
These temperatures are consistent with
our spectral fits of CI Tau in high state (Table 2).

Light curve shapes were classified by Wolk et al. (2005) as 
(i) linear rise with exponential decay, including rapid onset
($\ltsimeq$1 hr) impulsive flares,
(ii) nearly symmetric rise and decay, and (iii) short duration spikes ($<$5 ks).
But low-mass pre-main sequence stars show a wide variety of X-ray
light curve flare shapes including rapid impulsive rises followed
by slower rising secondary peaks during decay indicative of reheating events
(e.g. the $\rho$ Oph source GY 195 shown in Fig. 5 of 
Gagn\'{e}, Skinner, \& Daniel 2004). Since the Chandra observations
of CI Tau reported herein did not fully capture the rise and
decay phases during episodes of variability the overall light
curve shapes are not completely determined. But the slow rise times of
several hours rule out a spike. The nonlinear count rate increase 
and slow turnover during ObsId 29122 (Fig.1-bottom) 
suggests that it may be of symmetric type but this is 
not assured in the absence of observational constraints 
on the decay shape and timescale.

Even though the count rate, flux, and emission measure
were increasing during the two high state observations the 
time-partitioned data show little if any significant change
in the hardness ratio or plasma temperature. This is a clear
indication that the variability is a consequence of an
increase in the amount (volume emission measure) of hot plasma 
at a nearly-constant temperature rotating into the line of sight. 
Since the count rate must have peaked at least twice during the
$\approx$22 hr time interval spanned by ObsIds 28501 and 29122
(i.e. once in the gap between the observations and again at
the end of ObsId 29122), the variability may have
originated from one or more relatively small-scale  
inhomogeneous structures.

Rotation of long-lived ($>$P$_{rot}$) active regions across
the line-of-sight can give rise to phase-locked X-ray
variability in young stars (Flaccomio et al. 2005). The relative 
amplitudes determined for the sample of modulated Chandra COUP sources with
known optical periods identified by Flaccomio et al.
are Amp$_{\rm rel}$ = 0.18 - 0.72 as determined using the minimum and 
maximum count rates  Amp$_{\rm rel}$ = (max$-$min)/max$+$min).
For CI Tau we take min = 7.82 c ks$^{-1}$ from the two low-state
observations. The maximum count rate for both high state observations
occurred in the third segment (t3) and was max $\geq$ 44.9 c ks$^{-1}$
for ObsId 28501 since the rate was still increasing and
max = 51.1  c ks$^{-1}$ for ObsId 29122. These values give
Amp$_{\rm rel}$ $\geq$ 0.70 (ObsId 28501) and
Amp$_{\rm rel}$ = 0.73 (ObsId 29122), both at the high end of
the range of the COUP sample.

The modulated segments in the light curves shown in Flaccomio et al.
generally have rise times from nominal to peak count rate 
spanning at least $\Delta$$\phi$ = 0.1 - 0.2 in rotational phase. By comparison, 
the inferred rise time to peak count rate for CI Tau in ObsId 29122 is 
$\approx$4 hours, or equivalently $\Delta$$\phi$ = 0.02 assuming
P$_{rot}$ = 9 days, much faster than observed in the modulated COUP sample.
The high amplitude of CI Tau's X-ray variability and abrupt rise 
as measured by $\Delta$$\phi$ point to a relatively compact  
X-ray bright structure on the star rotating into the line-of-sight.
 
Finally, we note that a coherent period of 14.2 d was reported 
for CI Tau by Manick et al. (2024).
The elapsed time between ObsId 29094 and ObsId 29122 spans 14.2 d but X-ray 
variability was only detected in the latter. So there is no indication of
a 14.2 d X-ray modulation. X-ray monitoring of CI Tau to date
is quite limited and spans only a small fraction of the star's suspected
9 day rotation period. Additional monitoring over multiple stellar 
rotations would be needed to determine if any phase-locked 
rotational X-ray modulation is present.

\section{X-ray Irradiation of the Planet}

\subsection{Incident X-ray Flux}

The detection of strong X-ray variability in CI Tau implies that the X-ray
flux incident on the claimed planet is also 
variable. Assuming X-ray absorption
in the region between the star and planet is negligible, the unabsorbed flux 
incident on the planet at a distance r$_{\rm au}$ (au) from the star is
F$_{x,unabs}$ (erg cm$^{-2}$ s$^{-1}$)
= 3.557 $\times$ 10$^{-28}$ L$_{x}$(erg s$^{-1}$)/r$_{\rm au}^2$.
As determined in Section 3.2 we use characteristic X-ray luminosities 
log L$_{x}$ (erg s$^{-1}$) = 29.74 (low state) and 30.50 (high state).
The unattenuated flux at the planet is then
F$_{x,unabs}$ (erg cm$^{-2}$ s$^{-1}$) = 195.5/r$_{\rm au}^2$ (low state)
and 1124.8/r$_{\rm au}^2$ (high state).
Taking r$_{\rm au}$ = 0.17 (Manick et al. 2024) for the planet gives
F$_{x,unabs}$ (erg cm$^{-2}$ s$^{-1}$) = 6.77 $\times$ 10$^{3}$  (low state)
and 3.89 $\times$ 10$^{4}$ (high state).
For comparison, these fluxes are $\sim$10$^{5}$ - 10$^{6}$ times greater
than the Sun's X-ray flux at Jupiter assuming a nominal solar
X-ray luminosity log L$_{x,\odot}$ = 27.3 erg s$^{-1}$.

\subsection{X-ray Ionization and Heating of the Planet's Atmosphere}

To estimate the planet's ionization and heating rates
we follow the procedure in Skinner \& G\"{u}del (2024; hereafter SG24) and
references therein (e.g. Shang et al. 2002), as summarized below.

The X-ray ionization rate $\zeta(r)$ 
per H nucleus in the planet's atmosphere evaluated at a distance $r$
from the star for a thermal plasma at temperature kT is (eq. [1] of SG24)
$\zeta \approx \zeta_{\rm x}(r/\rm {R_{x}})^{-2}(\rm {kT}/\epsilon_{ion}) {\rm e}^{-\tau_{\rm x}} \ 
({\rm s}^{-1}~{\rm per~ H~nucleus)}$
where $\zeta_{\rm x}$ is the primary ionization rate defined below, 
$\epsilon_{ion}$ $\approx$ 37 eV  is the energy to create an
ion pair, R$_{x}$ is the radius at which the X-ray emission originates, 
and e$^{-\tau_{x}}$ is the X-ray attenuation due to atmospheric absorption
evaluated at optical depth $\tau_{x}$. The total ionization rate above
accounts for multiple secondary ionizations spawned by the primary ionization $\zeta_{\rm x}$.
As in SG24 we assume 
R$_{x}$ $\approx$ R$_{*}$ where the stellar radius is
R$_{*}$ = 2 R$_{\odot}$ (Donati et al. 2020). Inserting this value of
$R_{x}$ into eq. (2) of SG24 
and evaluating at the planet's nominal separation $r$ = 0.17 au = 18.3 R$_{*}$
gives~
$\zeta_{\rm x} =  5.85 \times 10^{-7}(L_{x}/10^{30}~ {\rm erg~ s}^{-1})(\rm{kT/keV})^{-(p+1)}$ (s$^{-1}$).
The value of $p$ is abundance dependent and $p$ = 2.485 for solar 
abundances (Morrison \& McCammon 1983; Shang et al. 2002),
which we adopt here. If the X-rays originate above the star then 
R$_{x}$ $>$ R$_{*}$ and the above value of $\zeta_{\rm x}$ is a lower limit.
  
In low state when the emission of CI Tau is steady the X-ray spectral fits
give log L$_{x}$ = 29.74 erg s$^{-1}$ and  kT $\approx$ 2 keV (Table 2). 
The low state primary ionization rate is thus
$\zeta_{\rm x}$ = 2.86 $\times$ 10$^{-8}$ s$^{-1}$. 
Inserting this value into eq. (1) and evaluating at the 
planet's separation 0.17 au gives
$\zeta$ = 4.6 $\times$ 10$^{-9}$e$^{-\tau_{x}}$ (s$^{-1}$ H$^{-1}$).
Evaluating at X-ray optical depth unity ($\tau_{x}$ = 1) in the planet's 
atmosphere leads to the result
$\zeta$ = 1.7 $\times$ 10$^{-9}$ (s$^{-1}$ H$^{-1}$). 

In high state the emission is variable but to obtain an 
estimate we adopt representative values 
log L$_{x}$ $\approx$ 30.5 ergs s$^{-1}$ and 
kT $\approx$ 4 keV (Table 2). These  values give
an ionization rate $\zeta$ that is nearly identical to low state.
This is a consequence of the steep inverse 
temperature dependence (kT)$^{-(p+1)}$ in $\zeta_{\rm x}$.
At higher X-ray temperatures the X-ray photons have higher
average energies and penetrate deeper in the atmosphere.
They are less absorbed as a result of the decrease in 
X-ray photoelectric absorption cross section with 
energy $\sigma$(E) $\propto$ E$^{-p}$ (Morrison \& McCammon 1983).

The X-ray heating rate per unit volume is
$\Gamma_{\rm x}$ = $\epsilon_{x}\zeta n_{\rm H} Q$
where 0 $<$ $\epsilon_{x}$ $<$ 1 is the fractional X-ray heating efficiency,
$Q$ $\approx$ 20 eV is the heating rate per ionization, and
$n_{\rm H}$ is the number density of hydrogen nuclei in the planet's
atmosphere at the height corresponding to the X-ray optical depth $\tau_{x}$
at which $\zeta$ is computed. A model of the planet's atmosphere
is needed to obtain the run of $n_{\rm H}$ versus depth. Using the above low state
value of $\zeta$ at $\tau_{x}$ = 1 gives 
$\Gamma_{\rm x}$ = 5.4 $\times$ 10$^{-20}$$\epsilon_{x}$ (erg cm$^{-3}$ s$^{-1}$ $n_{\rm H}^{-1}$).

\subsection{Planetary Mass Loss}

A reliable determination of the planet's photoevaporative mass loss rate 
requires knowledge of the star's unattenuated X-ray and EUV (XUV)
fluxes or luminosities and detailed hydrodynamic
modeling of the planet's atmosphere. Except for a few nearby stars,
EUV emission (0.013 - 0.1 keV) cannot be directly observed due to
interstellar absorption and must be estimated. The approximation
L$_{\rm EUV}$ $\sim$ L$_{x}$ has been used for 
late-type stars (e.g. Ribas et al. 2005; Owen \& Jackson 2012)
and is probably accurate to within a factor of a 
few for TTS at the $\sim$2 Myr age of CI Tau (Tu et al. 2015).
Observational constraints on the atmospheric properties of CI Tau c 
needed to guide hydrodynamic models
are lacking. However, an order-of-magnitude estimate of the 
planet's photoevaporative mass loss
rate can be obtained using the energy-limited approximation, as
discussed for TAP 26 b by SG24 and summarized below.

The energy-limited mass loss rate is (Erkaev et al. 2007; Sanz-Forcada et al. 2011)
$\dot{M}_{p,el}$ = 3F$_{\rm XUV}$/(4G$\rho_{p}\mathcal{K})$ where
$\rho_{p}$ is the planet's mean mass density, 0 $<$ $\mathcal{K}$ $\leq$ 1
accounts for Roche lobe effects, G is the gravitational constant and
F$_{\rm XUV}$ is the unattenuated XUV flux at the planet's surface.
Normalizing to Jupiter's mass density $\rho_{J}$ = 1.3 g cm$^{-3}$,
assuming negligible Roche lobe effects ($\mathcal{K}$ $\approx$ 1; 
Erkaev et al. 2007; Lammer et al. 2009),
and expressing the unattenuated F$_{\rm XUV}$ flux in terms of
XUV luminosity L$_{\rm XUV}$ = L$_{x}$ $+$ L$_{\rm EUV}$ leads to
$\dot{M}_{p,el} \approx 5\times10^{-12}(\rho_{p}/\rho_{\rm J})^{-1}(a/{\rm 0.1~ au})^{-2}(L_{\rm XUV}/10^{30}~ {\rm erg~ s^{-1}})~~M_{\rm Jup}~{\rm yr}^{-1}$.
Using the approximation L$_{x}$ $\sim$ L$_{\rm EUV}$ as mentioned above
and considering both the low and high state L$_{x}$ values (Sec. 3.2) for CI Tau 
and assuming the planet's nominal separation of 0.17 au yields
$\dot{M}_{p,el}$ $\sim$ 2 $\times$ 10$^{-12}$ $M_{\rm Jup}~{\rm yr}^{-1}$ (low state)
and $\sim$ 1 $\times$10$^{-11}$ $M_{\rm Jup}~{\rm yr}^{-1}$ (high state).
These energy-limited mass loss rates are order-of-magnitude estimates and 
subject to refinement by more detailed modeling of the planet's atmosphere
but the rates are quite low and $\dot{M}_{p}$ will generally decrease with time as CI Tau
ages and its XUV emission declines.

\subsection{Implications of the Planet's Eccentric Orbit}

The above calculations assume the putative planet's separation to be the nominal
value 0.17 au = 18.3 R$_{*}$. However, the planet's orbit is elliptical with 
eccentricity $e$ = 0.58$^{+0.05}_{-0.06}$ (Manick et al. 2024).
Thus, the separation will vary during the orbit over
the range $r_{min,max}$ = 0.17(1 $\pm$ e), or 
$r_{min}$ = 0.07 au = 7.5 R$_{*}$ and $r_{max}$ = 0.27 au = 29 R$_{*}$.  
The X-ray ionization rate is proportional to $r^{-2}$ and so is
the energy-limited mass-loss rate through its dependence 
on F$_{\rm XUV}$. As such, the nominal values of $\zeta$ and $\dot{M}_{p,el}$
computed above will increase by a factor of 5.9 near periastron and
decrease by a factor of 2.5 near apastron. Although the photoevaporative
mass-loss rate is higher near periastron, a larger fraction of the 
planet's orbital period is spent further out near apastron as
required by Kepler's second law. The elliptical orbit may lead to
other variable effects such as periodic pulsed accretion onto the star
as discussed by Biddle et al. (2018), Teyssandier \& Lai (2020),
and Manick et al. (2024), but Donati et al. (2024) concluded that
the accretion is stable.

\section{Summary}

The new Chandra observations of CI Tau confirm that its X-ray emission is variable.
The average plasma temperature and luminosity ranged from
kT $\approx$ 2 keV and log L$_{x}$ = 29.74 erg s$^{-1}$ during steady 
emission to kT $\approx$ 4 - 5 keV and log L$_{x}$ = 30.5 erg s$^{-1}$ 
in high state when the count rate was increasing, thus reaching an X-ray luminosity
$\approx$5 times higher than recorded in a 2005 XMM-Newton observation.
The observed high state kT and  L$_{x}$ values are comparable to other
young solar-like stars of mid-K spectral type during periods of elevated emission.
Very few clearly identifiable emission lines are seen in the X-ray
spectra but during periods of higher count rate and temperature
the emission near 1.86 keV (Si XIII) was enhanced.
A complete picture of the 
light curve variability profile is lacking due to gaps between
Chandra observations 
but the inferred rise time of the count rate
in the final Chandra observation (ObsId 29122) is $\approx$4 hours,
a small fraction $\Delta$$\phi$ = 0.02 in rotational phase assuming
a 9 day rotation period.
During periods of variability when the count rate, flux, and emission 
measure were increasing the hardness ratio and mean plasma temperature
remained nearly steady.  This is consistent with X-ray variability 
driven by an increase in the amount (volume emission measure)  
of hot plasma  rotating into the line-of-sight and light curve
analysis suggests a relatively compact inhomogenous emitting region.
Additional X-ray monitoring of CI Tau is needed to more completely 
characterize the light curve morphology and variability timescales 
and search for rotational modulation.
The incident X-ray flux at the putative planet is
$\sim$10$^{5}$ - 10$^{6}$ times greater than the Sun's X-ray flux at Jupiter.
The energy-limited approximation leads to an order-of-magnitude estimate of
the planet's photoevaporative mass-loss rate 
$\dot{M}_{p,el}$ $\sim$ 10$^{-12}$ - 10$^{-11}$ $M_{\rm Jup}~{\rm yr}^{-1}$,
assuming the (unobserved) stellar EUV luminosity is comparable to the
X-ray luminosity.  
 
\begin{acknowledgments}
Support for this work was provided by  {\em Chandra} award
number GO4-25004X issued by the {\em Chandra} X-ray Center, which is operated by
the Smithsonian Astrophysical Observatory (SAO) for and on behalf of NASA.

\noindent This paper employs a list of {\em Chandra} datasets obtained by the
Chandra X-ray Observatory contained in 
~\dataset[DOI:10.25574/cdc.473]{https://doi.org/10.25574/cdc.473}.
\end{acknowledgments}

{\em Facilities:} CXO
\software{XSPEC (Arnaud 1996),
          CIAO (Fruscione et al. 2006)}

\clearpage

\end{document}